# Using Randomness to decide among Locality, Realism and Ergodicity.

Alejandro Hnilo
*CEILAP, Centro de Investigaciones en Láseres y Aplicaciones, (MINDEF-CONICET);*
*J.B. de La Salle 4397, (1603) Villa Martelli, Argentina.*
*email: alex.hnilo@gmail.com*
March 7th, 2021.

Loophole-free experiments have demonstrated that at least one of three features is false when the violation of Bell's inequalities is observed: Locality, Realism or (what is lesser known) Ergodicity. An experiment is proposed to find out, or at least to get an indication about, which one is false. It is based on recording the time evolution of the rate of series of outcomes that are found not-random in a pulsed Bell's setup. The results of such experiment would be important not only to the foundations of Quantum Mechanics. For, even if the foundational issue remained not fully decided, they would have immediate practical impact on the efficient use of quantum-certified Random Number Generators and the security of Quantum Key Distribution using entangled states.

**1. Introduction.**

It is well known that Bell's inequalities reveal the predictions of Quantum Mechanics (QM) to be incompatible with the intuitive features of Locality (L) and Realism (R) [1]. A bunch of experiments testing Bell's inequalities [2-7] practically closed all known technical imperfections, reaching the *loophole-free* condition [8-12]. QM predictions were confirmed.

Yet, at least one feature additional to L and R must be assumed *in the experiments*. It is the validity of the ergodic hypothesis at the hidden variables level. This was demonstrated by V.Buonomano in 1978 [13], and independently rediscovered with different names [14-17]. In short, assuming Ergodicity valid is necessary to insert data recorded with different experimental settings (and, unavoidably, at different times) into a single expression (the Bell's inequality). The data are time averages. Bell's inequality is made of integrals over the hidden variables (ensemble averages). Ergodicity (E) means time and ensemble averages are equal, and hence, that the latter can be correctly replaced by the former in that expression.

I warn the Reader that there are many definitions of R. The issue is complex and subtle [18]. What is assumed in this paper is R as defined *in the framework of the experimental violation of Bell's inequalities* [1]. I also warn that the term "non-Local" is often used in the literature to mean "non-Classical". Yet, quantum non-Locality is not a *fact* [19-21] but a hypothesis, on an equal footing with non-Realism and non-Ergodicity. Precisely, the aim of this paper is to propose an experiment to put some light on this issue.

So, loophole-free experiments demonstrate that descriptions of the observed violation of Bell's inequalities cannot simultaneously have the features of being L, R and E. Figure 1 is a Venn's diagram that represents these features in the space of the theories that can describe Bell's experiment. For example, the Copenhagen and Everett's many-worlds interpretations of QM, and extended probabilities [22] do not hold to R, and are hence placed outside the red set in this diagram. Bohmian Mechanics and Wheeler-Feynmann theory of radiation [23] do not hold to L, and are hence outside the blue set. Non-stochastic models [17], singular distributions [24] and non-homogeneous dynamics [14] do not hold to E, and are hence outside the yellow set. The list is not complete; the mentioned theories are just examples of elements in that space. Theories inside the black "triangle" with red, blue and yellow sides, in the intersection of the three main sets, are not mentioned here for they have been already refuted by the loophole-free experiments.

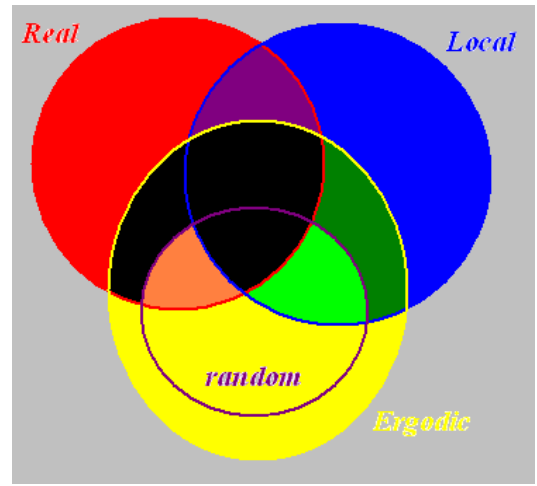

Figure 1: Sets in the space of the theories that describe Bell's experiment. Theories having the properties of being Real (red) *and* Local (blue) *and* Ergodic (yellow) have been experimentally refuted, and are indicated in black. The ones that are not-Local and not-random (in the sense that they produce not-random series of outcomes, see Fig.2) are also refuted, for they would allow faster-than-light signaling. Not-Local (but random and Real) theories are in orange; they support Quantum Certified Randomness. Not-Real (but Local and Ergodic) theories are in green (both light and dark) and include the Copenhagen interpretation of QM. Not-Ergodic (but Real and Local) theories are in violet, they produce not-random series of outcomes.

A relevant question now is: *which one of the three features (L,R or E) is false when Bell's inequalities are experimentally violated?* Of course, more than one can be false. I consider here the cases where only one of them is false. The aim of this paper is to propose an experiment to reveal (or, at least, to get some evidence to indicate) that false feature. Answering this question is of obvious importance for the foundations of QM, and also has practical consequences.

The key to the answer is the relationship between *falsity* of one of the three features and *randomness* of the series of outcomes produced in the Bell's setup (see Figure 2). As it will be explained, non-Locality implies the series are random (what is well known), non-Ergodicity implies the series are not-random, and non-Realism leaves the series' randomness undecided. Then, an experiment measuring the series' randomness would be able to answer the question. But measuring randomness, even *defining* randomness, is not easy. A given series cannot be demonstrated random. Yet, it can be demonstrated *not*-random. This occurs if it is rejected by one of the many existing tests of randomness, both statistical and algorithmic. The rate of rejected series (in a large set of series recorded under identical conditions) is a usual way to evaluate the reliability of random number generators (RNG) codes or devices in practice.

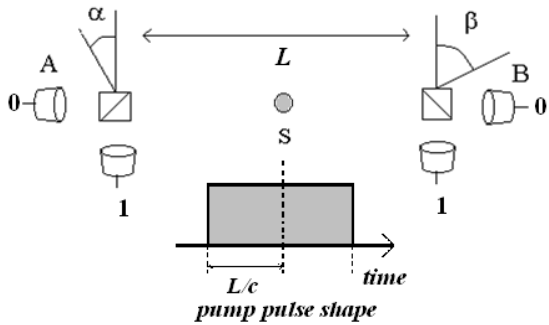

Figure 2: Sketch of the proposed experiment (Section 3.2). Source S emits pairs of photons maximally entangled in polarization during pulses of duration $\approx 2L/c$, analyzers are set at angles $\alpha,\beta$ at stations A and B placed at distance $L$; single-photon detectors are placed at the output gates of the analyzers, producing binary sequences. Settings $\alpha,\beta$ are changed between pulses, but remain fixed during the pulses. The loophole-free condition is enforced during the first part of the pulses ($t< L/c$), but not during the second part.

In consequence, I propose here to evaluate the level of randomness by measuring that rejection rate. Note that this is different from the usual way to evaluate randomness when quantum devices are involved, which is derived from assuming the existence of quantum non-Local effects (see Section 2.1). The reason to propose a different way is that *nothing* can be assumed here about the validity of L,R or E. Otherwise, one would fall into a logical inconsistency.

The rejection rate does not measure randomness, but I find hard to believe that it has *no relationship at all* with "actual randomness" (no matter how it is defined). The existence of *some* relationship is all what is needed here (see Section 3.1). Moreover, measuring the rejection rate immediately provides handy advices for the efficient use of RNG based on quantum measurements [25,26] and the secure use of Quantum Key Distribution setups using entangled states (QKD) [27] (see Section 3.3).

In the next Section 2, the consequences of the non-validity of L,R, or E are discussed. In the Section 3, the proposed experiment is described.

## 2. Consequences of non-validity of L,R or E.
### 2.1 Non-Locality and randomness.

"Random" is a property difficult to define, so that the idea of *quantum certified randomness* (QCR) is most welcome. According to QCR, the binary series of outcomes produced in the setup of Fig.2 are *intrinsically* random (in ideal conditions, of course). Such setup would provide then not only series to be used in practice, but also a precise definition: *random series is what is produced by this setup*. The level of randomness is proven to be directly related with the level of entanglement [25,26]. It is also shown that perfect random bits can be generated from imperfect series [28], what is of evident practical importance.

The idea and methods of QCR are developed from the following observation: *if* quantum non-Locality is taken *as an axiom*, then the series of outcomes of measurements performed on a (spatially spread) entangled state *must* be unpredictable. Otherwise, faster than light signaling would be possible [29]. In short: non-Locality $\Rightarrow$ random series.

However, in this paper, non-Locality is not an axiom, but a feature which validity is to be determined. Non-Locality is here on an equal footing with the alternatives of non-Realism and non-Ergodicity. If the methods related with QCR were used here, non-Locality would be implicitly assumed. I stress I do not mean these methods are erroneous. I just point out that, in this paper, they cannot be used without falling into a logical inconsistency. Therefore, here is forbidden (so to say) using the ideas and methods of QCR. On the other hand, this is one of the reasons why the experiment to be proposed is important: its results may support the hypothesis of quantum non-Locality, and hence, QCR.

### 2.2 Non-Ergodicity and randomness.

In general, the ergodic hypothesis means that the average of the variables of a dynamical system calculated over the phase space (*ensemble* average) is equal to the average obtained over the actual evolution of the system (*time* average). An ergodic evolution fills the system's phase space in a uniform way. If a system has a not-ergodic evolution instead, then it spends more time in some regions of phase space than in others. In consequence, it is more probable to find it in some regions than in others, and some prediction about its state can be done. Therefore, an evolution that is not ergodic $\Rightarrow$ it is (at least partially) predictable $\Rightarrow$ it is not random, for all conceivable definitions of "random". Be aware that this is valid for a system with a bounded phase space (or for a finite series, see below). The unbounded random walk is not ergodic.

In turn, *uniformity* of a binary series means that the rate of strings of length $n$ (say, 110100 for $n=6$) in the series is the same that would be obtained by tossing an ideal coin. Uniformity is considered the weakest form of randomness [30]. Let assume that the binary series produced in the setup of Fig.2 are the result of the evolution of an underlying dynamical system. Then each of those strings corresponds to a region in the phase space of that system. If the evolution is ergodic, then the system visits these regions equally. In consequence, the corresponding strings appear equally and the series is uniform. On the contrary, a system with a non-ergodic evolution visits the system's phase space unequally, and the resulting series is not uniform. As uniformity is the weakest form of randomness, a non ergodic evolution does not produce random series, even if no unanimously accepted definition of "random" exists. In short, then: non-ergodicity $\Rightarrow$ non-random series. By logical inversion: random series $\Rightarrow$ ergodicity valid.

Drawing the last implication into the diagram in Fig.1 leads to interesting conclusions. As noted in Section 2.1, not-Local theories that produce not-random series allow faster-than-light signaling and must be rejected. This is the black area additional to the "triangle" mentioned before. The theories in the orange painted set are not-Local, Real, Ergodic and produce random series. This is, precisely, the realm of QCR. On the other hand, not-Ergodic theories (violet region) produce not-random series.

*2.3 Non-Realism and randomness.*

Not-Real (or counterfactual indefinite [1,18]) but L and E theories are represented in Fig.1 by the green areas. The most important, by far, in this set of theories is the Copenhagen interpretation of QM. Strictly speaking, the Copenhagen interpretation says *nothing* about randomness of a series of outcomes of successive measurements made on a set of identically prepared states. Born's rule allows calculating probabilities, but not the features of the time series that lie beneath the measurement of such probabilities.

The only explicit definition on this subject is von Neumann's axiom, which states that quantum measurements violate Leibniz's principle of sufficient reason. In other words: a quantum measurement produces one or another outcome *without cause*. A series of such outcomes is intuitively random, although this intuition is difficult to formalize [31]. Besides, von Neumann's axiom can be understood in two ways, or strengths. Its "strong" form means that Leibniz's principle is violated at the quantum scale. The "weak" form means that the axiom is part of a user's guide or warning about what QM can or cannot calculate, but not necessarily a feature to be observed in the experiments.

Therefore, it is not known for sure if (always according to the Copenhagen interpretation, which assumes non-Realism valid) the series of outcomes are expected to be random (light green set in Fig.1) or not (dark green set).

**3. Experiment proposed.**
*3.1 A previous experimental result.*

An important experimental result is that the rejection rate reported by existing quantum-RNG, including Bell's setups, is about 30% [19,32-36]. This somehow disappointing performance is explained by taking into account that, even assuming non-Locality valid, the loophole-free condition is mandatory to ensure QCR [31]. The reported devices operate in a condition which is, at best, mixed between loophole-free and not-loophole-free. There is no reason to expect the setup to be proposed to show an *average* rejection rate much different than that number (this can be experimentally checked anyway). As it is explained next, the observed intermediate value ($\approx 30\%$) of the rejection rate opens the door to an experiment to get some indication about the false feature, by using simple numerical properties only.

*3.2 Basic scheme.*

Consider the setup in Fig.2. Assume the source emits maximally entangled states during square pulses with short risetime and total duration about twice longer than $L/c$, where $L$ is the distance between stations and $c$ is the speed of light. This is achieved by using a pulsed laser pump. Angle settings $\{\alpha,\beta\}$ are changed between pulses, but they are left static during the pulses. If a detection occurs at the transmitted (reflected) port, a "0" ("1") is written in a recording device together with the time value it occurred (*time stamping*). Samples of the laser pump pulses are sent to each station to synchronize the clocks. Data processing identifies the coincidences, singles are discarded. Time between pulses is much longer than their duration. After many pulses, binary sequences of coincidences are recorded.

Assume now that the violation of Bell's inequalities is constant throughout the pulse duration, as predicted by QM and also experimentally reported [5,33]. During the pulses' first part the detections are spatially isolated. If the detectors are efficient enough, the loophole-free condition is enforced. Therefore, during this period the violation of Bell's inequalities is possible only because L, *or* R, *or* E is false. During the pulses' second part instead, there has been enough time for the stations to interchange information (recall analyzers' settings are fixed during the pulse), the loophole-free condition is no longer valid, and Bell's inequalities can be violated even if the three features are all true.

The produced series can be then divided in two sets: the ones made of coincidences recorded during the pulses' first part, when the loophole-free condition is enforced, and the ones made of coincidences recorded during the second part, when it is not.

If the violation of Bell's inequalities observed during the first part (loophole-free) of the pulses is possible because L is false (but not R, E), then we are in the orange set in Fig.1, the realm of QCR. Series recorded during this period *must* be truly random. Unless the rejecting rate and "actual randomness" are fully uncorrelated (what is improbable), tests of randomness applied to a set of these series should reject almost no series at all (= negligible rejection rate, say, $\approx 5\%$, not 0% because of some remaining technical imperfection). Instead, the series recorded during the pulses' second part (not loophole-free) may be or may be not random. As said, the average between the rejection rate values of these two sets of series should be similar to the reported intermediate value (≈30%). Therefore, the rejection rate for series recorded during the pulses' second part must *increase* in order to balance, in the average, the negligible rejection rate of the series recorded during the pulses' first part.

If the violation of the Bell's inequalities observed during the pulses' first part is possible because E is false instead (but not R, L), then we are in the violet set in Fig.1. Series recorded during the pulses' first part (loophole-free) *must* be not-random. It is then to be expected that most of them are rejected by the tests (= maximal rejection rate, say, ≈95%, not 100% because of some imperfection in the tests). Instead, the series recorded during the second part (not-loophole-free) may be or may be not random. The average between the rejection rate values of these two sets of series must be similar to the reported intermediate value (≈30%). Therefore, the rejection rate for series recorded during the pulses' second part must *decrease* in order to balance, in the average, the high rejection rate of the series recorded during the pulses' first part.

Finally, if the violation of the Bell's inequalities observed during the pulses' first part is possible because R is false instead (but not L, E), then we are in the green set in Fig.1, the realm of the Copenhagen interpretation. The series may be random (light green) or not (dark green), but this is so regardless they are recorded in the first or the second part of the pulses. In the average over many experimental runs, the rejection rate should be observed nearly constant in time.

Usual technical imperfections are commented in Section 3.4. At this point, it suffices to say there is no reason for them to affect the first or the second pulses' parts in a different way. Hence, they may affect the average value of the total rejection rate, but not its time evolution.

In order to estimate the statistical deviation to confirm the observed trend, the rejection rate must be calculated for different choices of the sets of tests (see Section 3.3). If sufficient data are available, the pulses can be sliced in more than two parts and a *curve* of evolution of the rejection rate be plotted, that would be most desirable.

In summary: the consistent observation of an increasing (decreasing) trend of the rejection rate suggests L (E) is violated. A constant rejection rate suggests R is violated instead. Conceivably, the last result might also be caused by a high level of noise masking the actual trend. In the case the rejection rate is in fact observed to be constant within the statistical deviation, this alternative should be carefully examined by looking for sources of noise.

*3.3 Evaluating randomness.*

Entropies are often used to evaluate randomness, but they can only be properly calculated for stationary series of large length, which are difficult to obtain experimentally. Evaluating randomness using *only* entropies may force the experimenters to discard a large amount of data. Besides, a series can have maximal entropy and still be not-random, so that additional tests must be applied anyway.

The pragmatic approach proposed here is to evaluate randomness from the number of series (in a large ensemble recorded under the same conditions) rejected by a set of tests. In practice, a usual choice for the set is the NIST battery of 16 statistical tests.

Let suppose a set of series $S$ in which a subset $R$ are random. It is intuitive that the set of not-rejected series approaches to $R$ as the number and variety of the tests applied to $S$ increases. It seems then appropriate to make the set of tests as large as possible. F.ex., by adding estimators of Kolmogorov's complexity [37,38], that can be applied to non-stationary series. Also, tools of nonlinear analysis to identify a compact object in phase space (Takens' theorem) [39]. It is easy including the calculation of Hurst coefficient to measure self-correlation, and Augmented Dickey-Fuller (ADF) and Kwiatkowski-Phillips-Schmidt-Shin (KPSS) tests to check stationarity. If (when) the recorded series are stationary, the calculus of entropies can also be done. This list is just an example of the set of tests that can be used.

Evaluating randomness by using the rejection rate has the crucial advantage of making no assumption about the validity of L, R or E. On the other hand, the obtained value depends on the set of test chosen, which is arbitrary. For this reason, I claim the observation of a consistent trend in the time variation of the rejection rate to provide just *some evidence* of the falsity of L, R or E, not a *proof*.

In spite of this limitation, the results of the proposed experiment have an immediate practical impact, which is independent of the foundational issue. If the experiment showed the rejection rate to increase with time, then QKD using entangled states would be safer if pulses shorter than $L/c$ were used to generate the key. If it showed the rejection rate to decrease instead, the final part of long pulses (duration $> L/c$) should be preferred. Finally, if it showed the rejection rate to be constant, then both the pulse duration and the pulse's part used would be irrelevant (continuous-wave sources would be fine, too). Analogous advices would apply to the most efficient way (= lowest number of not-random series) to operate a quantum-RNG.

*3.4 Conditions for an attainable experiment.*

Unfortunately, the experiment as it is described in Section 3.2 is unattainable nowadays. Due to detectors' efficiencies, the loophole-free condition can be reached with photons only by using Eberhardt's states, which produce heavily biased series. Extraction methods can be applied to get uniform series [40,41], but their use in this case may mask the phenomenon one wants to detect. Setups exploiting entanglement swapping between photons and matter do use Bell states, but produce a rate of detections too low to be suitable.

A simple solution at hand is to accept the *fair sampling* assumption [1] valid. In this case, the set of recorded coincidences is assumed to be an unbiased statistical sample of the whole set of detected and not-detected biphotons. Under this assumption, Bell states and standard detectors can be used.

Another problem is achieving fast and random setting changes. In addition to the technical difficulty of fastness, there is the logical problem (a sort of infinite regress) of performing *random* setting changes. It has been demonstrated [28] that a small amount of initial randomness suffices to get fully random settings but, once again, this method cannot be applied here, for it is derived from assuming quantum non-Locality.

The problems of fastness and randomness can be circumvented by assuming that any hypothetical correlation between the stations vanishes when the source of entangled states is turned off. An observation supporting this assumption is that the curve of the values taken by the Clauser-Horne-Shimony-Holt parameter, as the time coincidence window is increased beyond the pulses' duration, decays following the curve predicted if the detections outside the pulse are uncorrelated [5,42]. Non-correlation implies the curve but, of course, observing the curve does not necessarily imply lack of correlation. Nevertheless, if the latter implication is accepted as true, then fast and random settings' changes become unnecessary. Only well separated pump pulses are required.

*Under these two assumptions* ("fair sampling" and, say, "uncorrelated when pump is turned off") the proposed experiment is at hand even with limited means. The results obtained in these conditions cannot be considered definitive, but they may still give a clue about the answer to the main question. They may also help to decide whether the effort to solve the problems of the complete experiment is worthy, or not. They would have a practical impact, too (see Summary).

Some numbers: in order to keep the rate of accidental coincidences low, source intensity must be adjusted such that the probability $p$ of detection per pulse is $p \ll 1$ [43]. Choosing $p=0.1$ and pulse repetition rate 1 MHz, sequences 6 Mbits long are recorded at each station in a run lasting 300s. It is not convenient increasing the repetition rate beyond that value, for $\approx 1$ MHz is the threshold of saturation of available single photon detectors (silicon avalanche photodiodes). If the stations are separated 20m, then the pulse duration is $\approx 120$ ns and the duty cycle is $\approx 12\%$. These numbers are easily achievable by pumping the nonlinear crystals that generate the entangled states with a pulsed diode laser, which typically has a TTL bandwidth of 20 Mhz.

Detectors' blind time was identified to cause non-uniform series in some quantum RNG. But this is not important in the proposed experiment, because the average number of detections per pulse is, as said, adjusted to be small ($p \ll 1$).

## Summary.

By recording the coarse time evolution of the rate of not-random series obtained in a suitable pulsed Bell's experiment, it is possible to get some evidence about which one among three fundamental features (Locality, Realism or Ergodicity) is false when Bell's inequalities are observed to be violated. This is of obvious interest for the foundations of QM.

An experiment achievable nowadays requires some additional assumptions, which weaken the consequences of its results from the foundational point of view. Nevertheless, even if the foundational issue remained not fully solved, the result of observing the variation of the rejection rate would have immediate impact on the best practical use of quantum-RNG and QKD with entangled states. Depending on said result, it may be advisable, in order to produce the smallest number of not-random series (in RNG) or to get a safe key (in QKD), using sources with pulses shorter than $L/c$, or instead using the end ($>L/c$) of long pulses, or it may also turn out that the pulses' duration is unimportant (see Section 3.3).

## Acknowledgments.

Many thanks to Marcelo Kovalsky for most useful discussions. This material is based upon research supported by, or in part by, the U.S. Office of Naval Research under award number N62909-18-1-2021. Also, by grant PIP 2017-027 CONICET (Argentina).

## References.

[1] J.Clauser and A.Shimony, "Bell's theorem: experimental tests and implications", *Rep. Prog. Phys.* **41** p.1881 (1978).
[2] A.Aspect *et al.*, "Experimental test of Bell's inequalities using time-varying analyzers", *Phys. Rev. Lett.* **49** p.1804 (1982).
[3] G.Weihs *et al.*, "Violation of Bell's inequality under strict Einstein locality conditions", *Phys. Rev. Lett.* **81** p.5039 (1998).
[4] M.Agüero *et al.*, "Time stamping in EPRB experiments: application on the test of non-ergodic theories"; *Eur. Phys. J. D* **55** p.705 (2009).
[5] M.Agüero *et al.*, "Time resolved measurement of the Bell's inequalities and the coincidence-loophole", *Phys. Rev. A* **86**, 052121 (2012).
[6] M.Giustina *et al.*, "Bell violation using entangled photons without the fair-sampling assumption", *Nature* **497** p.227 (2013).
[7] B.Christensen *et al.*, "Detection-loophole-free test of quantum nonlocality, and applications", *Phys. Rev. Lett* **111** 130406 (2013).
[8] M.Giustina *et al.*, "A Significant Loophole-Free Test of Bell's Theorem with Entangled Photons", *Phys. Rev. Lett.* **115**, 250401 (2015).
[9] L.Shalm *et al.*, "A Strong Loophole-Free Test of Local Realism", *Phys. Rev. Lett.* **115**, 250402 (2015).
[10] B.Hensen *et al.*, "Loophole-free Bell inequality violation using electron spins separated by 1.3 kilometers", *Nature* **526**, 682 (2015).
[11] W.Rosenfeld *et al.*, "Event-ready Bell-test using entangled atoms simultaneously closing detection and locality loopholes", *Phys. Rev. Lett.* **119**, 010402 (2017).
[12] A.Hnilo, "Consequences of recent loophole-free experiments on a relaxation of measurement independence"; *Phys. Rev. A* **95**, 022102 (2017).
[13] V.Buonomano, "A limitation on Bell's inequality", *Annales del'I.H.P. Sect.A*, **29** p.379 (1978).

[14] A.Hnilo, "Using measured values in Bell's inequalities entails at least one hypothesis additional to Local Realism", *Entropy* **19**,80 doi: 103390/e19040180 (2017).
[15] H.S.Poh *et al.*, "Probing the quantum–classical boundary with compression software", *New J.Phys.* **18** 035011 (2016).
[16] A.Khrennikov, "Buonomano against Bell: nonergodicity or nonlocality?", *Int. J. Quantum Inf.* **15** 1740010 (2017).
[17] K.Hess *et al.*, "Hidden assumptions in the derivation of the Theorem of Bell", *Phys. Scr.* 2012 01002.
[18] A.Cabello, "Interpretations of Quantum Theory: A Map of Madness", in: *What is Quantum Information?*,Cambridge University Press, Cambridge, UK, (2017) edited by O.Lombardi *et al.*, also *arXiv:1509.04711*.
[19] M.Kupczynski, "Closing the Door on Quantum Nonlocality", *Entropy* **2018**, 20, 877.
[20] A.Khrennikov, "Get Rid of Nonlocality from Quantum Physics", *Entropy* **2019**, 21, 806.
[21] A.Hnilo, "Quantum Mechanical description of Bell's experiment assumes Locality", *arxiv.org/2002.12153*.
[22] W.Mückenheim *et al.*, "A review of extended probabilities", *Phys.Reports* **133**(6) p.337 (1986).
[23] D.Pegg, "Time symmetric electrodynamics and the Kocher-Commins experiment"; *Eur.J.Phys.* **3** p.44 (1982).
[24] J.Geurdes, "A counter-example to Bell's theorem with a softened singularity and a critical remark to the implicit demand that physical signals may not travel faster than light", *arxiv:physics/0112036*.
[25] S.Pironio *et al.*, "Random numbers certified by Bell's theorem", *Nature* **23**, p.1021 (2010).
[26] N.Brunner *et al.*, "Bell nonlocality," *Rev.Mod.Phys.* **86**, p.419 (2014).
[27] A.Ekert, "Quantum cryptography based on Bell's theorem"; *Phys.Rev.Lett* **67**, p.661 (1991).
[28] R.Gallego *et al.*, "Full randomness from arbitrarily deterministic events", *Nat. Comm* **4**, 2654 (2013).
[29] S.Popescu and D.Rohrlich, "Quantum nonlocality as an axiom", *Found. Phys.* **24** p.379 (1994).
[30] G.Sommazzi, "Kolmogorov Randomness, Complexity and the Laws of Nature", https://www.researchgate.net/publication /311486382.
[31] A.Khrennikov, "Randomness: Quantum vs classical", *Int.J.of Quantum Inform.* **14**, 1640009 (2016).
[32] M.Kovalsky *et al.*, "Kolmogorov complexity of sequences of random numbers generated in Bell's experiments"; *Phys. Rev. A* **98**, 042131 (2018), see addendum on series of outcomes at: *arXiv.org/1812.05926*.
[33] M.Nonaka *et al.*, "Randomness of imperfectly entangled states"; *arXiv.org/abs/1908.10794*.
[34] C.Calude *et al.*, "Experimental evidence of quantum randomness incomputability", *Phys.Rev.A* **82**,022102 (2010).
[35] A.Solis *et al.*, "How random are random numbers generated using photons?", *Phys. Scr.* **90**, 074034 (2015).
[36] A.Hnilo et *al.*, "Low dimension dynamics in the EPRB experiment with random variable analyzers"; *Found.Phys.* **37** p.80 (2007).
[37] A.Kolmogorov, "Three approaches to the quantitative definition of information", *Problems of Information Transmission* **1** p.4 (1965).
[38] D.Mihailovic *et al.*, "Novel measures based on the Kolmogorov complexity for use in complex system behavior studies and time series analysis", *Open Phys.* 2015 13:1-14.
[39] H.Abarbanel, *Analysis of observed chaotic data* (Springer-Verlag, 1996).
[40] P.Bierhorst *et al.*, "Experimentally generated randomness certified by the impossibility of superluminal signals", *Nature* **556**, p.223 (2018).
[41] L.Shen *et al.*, "Randomness extraction from Bell violation with continuous parametric down conversion", *Phys. Rev. Lett.* **121**, 150402 (2018).
[42] M.Agüero *et al.*, "Time stamping in EPRB experiments: application on the test of non-ergodic theories", *Eur.Phys.J. D* **55** p.705 (2009).
[43] M.Agüero *et al.*, "Measuring the entanglement of photons produced by a nanosecond pulsed source", *J. Opt. Soc. Am. B* **31** p.3088 (2014).